\documentclass{emulateapj}

\usepackage{graphicx}
\usepackage{multirow}
\usepackage{color}
\usepackage{soul}
\usepackage{ot1patch}
\usepackage{color}
\usepackage[dvipsname2...................s]{xcolor}
\usepackage{verbatim}

\newcommand{\kyr}{\ensuremath{\,\mathrm{kyr}}}
\newcommand{\myr}{\ensuremath{\,\mathrm{Myr}}}
\newcommand{\gyr}{\ensuremath{\,\mathrm{Gyr}}}

\newcommand{\ma}{\ensuremath{M_{\rm a}}}
\newcommand{\mb}{\ensuremath{M_{\rm b}}}
\newcommand{\mzamsa}{\ensuremath{M_{\rm ZAMS, a}}}
\newcommand{\mzamsb}{\ensuremath{M_{\rm ZAMS, b}}}
\newcommand{\mzams}{\ensuremath{M_{\rm ZAMS}}}
\newcommand{\msun}{\ensuremath{\,M_\odot}}
\newcommand{\mmweg}{\ensuremath{\,M_{\rm MWEG}}}

\newcommand{\rsun}{\ensuremath{\,R_\odot}}

\newcommand{\zsun}{\ensuremath{Z_\odot}}

\newcommand{\startrack}{{\tt StarTrack}}

\newcommand{\dash}{\,\hbox{--}\,}

\newcommand{\ttt}[2]{\ensuremath{#1\times10^{#2}}}
\newcommand{\fqs}{\ensuremath{f_{\rm QS}}}
\newcommand{\fcr}{\ensuremath{f_{\rm cr}}}

\newcommand{\radir}{\ensuremath{\mathcal{R}_{\rm A,dir}}}
\newcommand{\raacc}{\ensuremath{\mathcal{R}_{\rm A,acc}}}
\newcommand{\azams}{\ensuremath{a_{\rm ZAMS}}}
\newcommand{\rbdir}{\ensuremath{\mathcal{R}_{\rm B,dir}}}
\newcommand{\rlmxb}{\ensuremath{\mathcal{R}_{\rm LMXB}}}
\newcommand{\rdqs}{\ensuremath{\mathcal{R}_{\rm DQS}}}
\newcommand{\nqs}{\ensuremath{\#\mathrm{QS}}}
\newcommand{\nns}{\ensuremath{\#\mathrm{NS}}}
\newcommand{\tage}{\ensuremath{t_{\rm age}}}
\newcommand{\mhmax}{\ensuremath{M_\mathrm{max}^H}}

\begin{document}

\title{Strange quark stars in binaries: \\formation rates, mergers and explosive phenomena}

\author{G. Wiktorowicz\altaffilmark{1},
        A. Drago\altaffilmark{2},
        G. Pagliara\altaffilmark{2},
        S.B. Popov\altaffilmark{3}} 

 \affil{$^{1}$ Astronomical Observatory, University of Warsaw, Al.  Ujazdowskie
 4, 00-478 Warsaw, Poland (\textit{gwiktoro@astrouw.edu.pl})\\
        $^{2}$ Dip.~di Fisica e Scienze della Terra dell'Universit\`a di Ferrara 
and INFN Sez.~di Ferrara, Via Saragat 1, I-44100 Ferrara, Italy\\
        $^{3}$Sternberg Astronomical Institute, Lomonosov Moscow State University, Universitetsky prospekt 13, 119234, Moscow, Russia }
 
\begin{abstract} 
Recently, the possible co-existence of a first family composed of "normal"
neutron stars with a second family of strange quark stars has been proposed as a
solution of problems related to the maximum mass and to the minimal radius of
these compact stellar objects.  In this paper we study the mass distribution of
compact objects formed in binary systems and the relative fractions of quark and
neutron stars in different subpopulations. We incorporate the strange quark star
formation model  provided by the two-families scenario and we perform a
large-scale population synthesis study in order to obtain the population
characteristics.  According to our results, the main channel for strange quark
star formation in binary systems is accretion from a secondary companion on a
neutron star. Therefore, a rather large number of strange quark stars form by
accretion in low-mass X-ray binaries and this opens the possibility of having
explosive GRB-like phenomena not related to supernovae and not due to the merger
of two neutron stars.  The number of double strange quark star's systems is
rather small with only a tiny fraction which merge within a Hubble time. This
drastically limits the flux of strangelets produced by the merger, which turns
out to be compatible with all limits stemming from Earth and Lunar experiments.
Moreover, this value of the flux rules out at least one relevant channel for the
transformation of all neutron stars into strange quark stars by strangelets'
absorption.

\end{abstract}


\section{Introduction}

The discovery, in 2010, of a pulsar with a mass of about two solar masses
\citep{Demorest1010} has stimulated many theoretical studies, in the nuclear
astrophysics community, concerning its possible composition and the properties
of the equation of state of dense matter. It is clear indeed that the center of
this stellar object could be the site of the most dense form of nuclear matter
we are aware of: depending on the adopted model for the equation of state, the
central density of this star could be larger than about 3 times the nuclear
saturation density. There are many different ideas on the composition of matter
at such a high density: for instance, hyperons \citep{Chatterjee:2015pua} or
delta resonances \citep{Drago:2014oja} could form, or a phase transition to
quark matter could occur \citep{Alford:2015dpa}. The need to fulfill the two
solar mass limit provides tight microphysical constraints on those scenarios.

It is clear that precise mass measurements represent a powerful and reliable
tool to investigate the properties of dense matter. Future observations (e.g. by
the FAST \citep{Nan11} and SKA radio telescopes \citep{Carilli:2004nx})
could possibly prove the existence of even larger masses. However, some
information could be obtained also by considering the (much more uncertain)
measurements of the radii.  Unfortunately, up to now, only in a few cases mass
and radius of a same compact star have been estimated from x-ray analysis and
moreover with large systematic uncertainties \citep{Ozel1609,Miller1603}.
Although still under debate, there are a few indications of the possible
existence of stellar objects with radii smaller than about 11~km
\citep{Guillot:2014lla,Ozel:2015fia}, thus very compact.  Very small radii for
stars having masses of about 1.4 -- 1.5 $M_\odot$ are obtained only if the
equation of state of dense matter is very soft at densities of about 2 -- 3
times nuclear matter saturation density. On the other hand, so soft equations of
state lead to maximum masses significantly smaller than $2\msun$, because to
reach very large masses would imply an extreme stiffening of the equation of
state at larger densities, saturating the limit of causality, a situation that
is not very realistic \citep{Alford:2015dpa}. If future observations with new
facilities, such as the NICER experiment on-board ISS \citep{Gendreau1209}, will
confirm the existence of very compact stars, then one has to explain how the
equation of state of dense matter could be at the same time very soft (to
explain the very compact configurations) and very stiff (to explain the very
massive configurations).  

In \citet{Drago:2013fsa,Drago1602,Drago:2015dea}, a possible solution to this
puzzle has been proposed. It is based on the existence of two families of
compact stars: neutron Stars (NSs indicating both stars made of nucleons and
stars containing hyperons) which are compact and light, and strange quark stars
(QSs) \citep{Alcock8611,Haensel:1986qb} which are large and massive (a $2\msun$
star would be a QS for instance). In this scenario, strange quark matter
composed of three flavors: up, down, and strange, is the true ground state and
hadronic matter is instead metastable.  A NS could therefore convert into a QS
once a significant fraction of strangeness is formed in its interior through the
appearance of hyperons and the conversion time turns out to be of the order of
ten seconds
\citep{Drago:2005yj,Herzog:2011sn,Niebergal:2010ds,Pagliara:2013tza,Drago:2015fpa}.
The critical density for such a transition is thus close to the threshold of
hyperons' formation \footnote{The critical fraction of hyperons needed to
    trigger the conversion can be estimated by requiring that the strange quark
    matter phase with the same strangeness fraction is energetically favored
    \cite{Iida:1998pi,Bombaci:2004mt}}. 
The exact value of the density at which hyperons start forming depends on the
microphysics of the equation of state and it determines the maximum mass and the
minimum radius of NSs (in this paper we do not consider rotating
configurations). The smaller the minimum radius the smaller the maximum mass.
Since we are interested in radii smaller than about 11~km, then the limiting
gravitational mass of a NS is $ M_\mathrm{max}^H \sim 1.5\dash1.6\msun$. As
shown in Fig.1, the unstable hadronic star forms a QS having the same baryonic,
but a smaller gravitational mass. The ``mass defect'' is $\Delta M \sim
0.1\dash0.15\msun$. One can infer therefore that within the two-families
scenario the mass distribution would be qualitatively different with respect to
the one of the standard one-family scenario: we expect, in particular, an
enhancement in the number of stars with masses in the range $(M_\mathrm{max}^H-
\Delta M) \leqslant M \leqslant M_\mathrm{max}^H$ (coexistence range),
compensated by a depopulation of the the region of masses larger than
$M_\mathrm{max}^H$, a feature that could be possibly tested by means of future
observations of the FAST and SKA radio telescopes.

We will focus here on binary systems with QSs, as they allow for the dynamical
mass measurement of a compact object. There exist three general ways a QS can
form during the evolution of a binary system\footnote{A further possibility is
    to form a strange quark star through the merger of two neutron stars, but
this scenario will not be analyzed in this paper.}:

\begin{enumerate}
    \item The binary components may not interact during their evolution
        (single-star-like evolution). It requires a star having an initial mass
        $M_{\rm ZAMS}\approx 18\dash22\msun$;
    
    \item The star may become a NS with a mass $M_{\rm NS}<M^H_{\rm max}$, and
        then accrete material from the companion to become a QS afterwards
        \citep[e.g.,][]{Cheng9608,Dai9811,Zhu1301,Jiang1507}. In this channel the
        initial mass range is much wider ($M_{\rm ZAMS}\approx 6\dash17.5\msun$)
        and more populated than in the previous case;  
    
    \item The third possibility involves mass-loss by a massive progenitor due
        to binary interactions: mass transfer (MT) or common envelope (CE). As a
        result, the mass of the post-supernova (SN) compact object will be lower
        than in the single-star evolution (for the same initial stellar mass).
        Instead of forming a black hole (BH), the star finishes its evolution
        producing a QS.  The pre-SN mass-loss is necessary for initial stellar
        masses $M_{ZAMS}\gtrsim22\msun$. It may be so prominent that the star
        will initially form a NS and transform into a QS only after a
        mass-accretion phase (see the second channel above).
 
\end{enumerate}

There are two particularly relevant results to be obtained: an estimate of the
number of QS-QS systems and of the number of low-mass X-ray binaries (LMXBs)
containing a QS.  The first number determines the number of double QS (DQS)
mergers (note that by DQS we denote only systems consisting of two QSs), which
in turn is related to the flux of strangelets ejected at the moment of the
merger.  Those strangelets can potentially trigger the conversion of all NSs
into QSs \citep{Madsen:1989pg,Madsen:2004vw} therefore invalidating the
two-families scenario.  We will address this problem in Sec \ref{sec:dqs}.  The
number of LMXBs containing a QS is potentially related to the number of (long)
$\gamma$-ray bursts generated by the exothermic NS to QS transition in a rapidly
rotating neutron star \citep{Drago:2015fpa}. This scenario could correspond to
GRB060614 in which a long $\gamma$-ray burst (GRB) was not accompanied by a
supernova \citep{Fynbo:2006mw,DellaValle:2006nf,GalYam:2006en}.

\section{Modeling}\label{sec:modeling}

We performed a simulation of 2 million binaries using the \startrack\ population
synthesis code \citep{Belczynski0206b,Belczynski0801} with some further
amendments \citep[see][and references therein]{Wiktorowicz1509}. These
large-scale simulation were obtained with a use of the Universe@Home
project\footnote{\it http://universeathome.pl}. Population synthesis method was
previously widely used to similar tasks \citep[e.g.][]{Popov0711}

We simulated a grid of six models with three different metallicities: \zsun\
\citep[solar metallicity; $\zsun=0.02$;][]{Villante1405}, \zsun/10, and
\zsun/100; and two values of $\mhmax$ parameter: $1.5$ and  $1.6\msun$.

For initial stellar mass distribution we used the \citet{Kroupa9306} broken
power-law with $\alpha=-2.3$ for stars heavier than $1\msun$.  For a primary we
chose a mass range of $6\dash150\msun$ to involve all possible progenitors of
compact objects. For secondaries we studied a wider range of $0.08\dash150\msun$
keeping the mass ratio distribution uniform ($P(q)=\mathrm{const}$). Initial
binary separations had the log-uniform distribution  --- so-called \"Opik law,
--- \citep[$P(a)\sim1/a$;][]{Abt83}, whereas, the eccentricity distribution was
assumed to be thermal \citep[$P(e)=2e$;][]{Duquennoy9108}. We assumed that the
natal kick acts only during NS formation (single-mode Maxwellian distribution
with $\sigma=265$ km s$^{-1}$).

\subsection{Strange quark star formation}

In this study every NS with a mass $M_{\rm NS}\geq M_{\rm max}^H$ transforms
into a QS.  The transition is so rapid that it occurs within a single time step
of our simulation \citep{Drago:2015fpa}.  In our results the maximum post-SN NS
mass was $1.924\msun$, which transforms into a $1.779\msun$ QS. (initially more
massive objects form BH). However, mass accretion may make QSs even heavier.

\begin{figure}[h]
    \centering
    \includegraphics[height=7cm,angle=0]{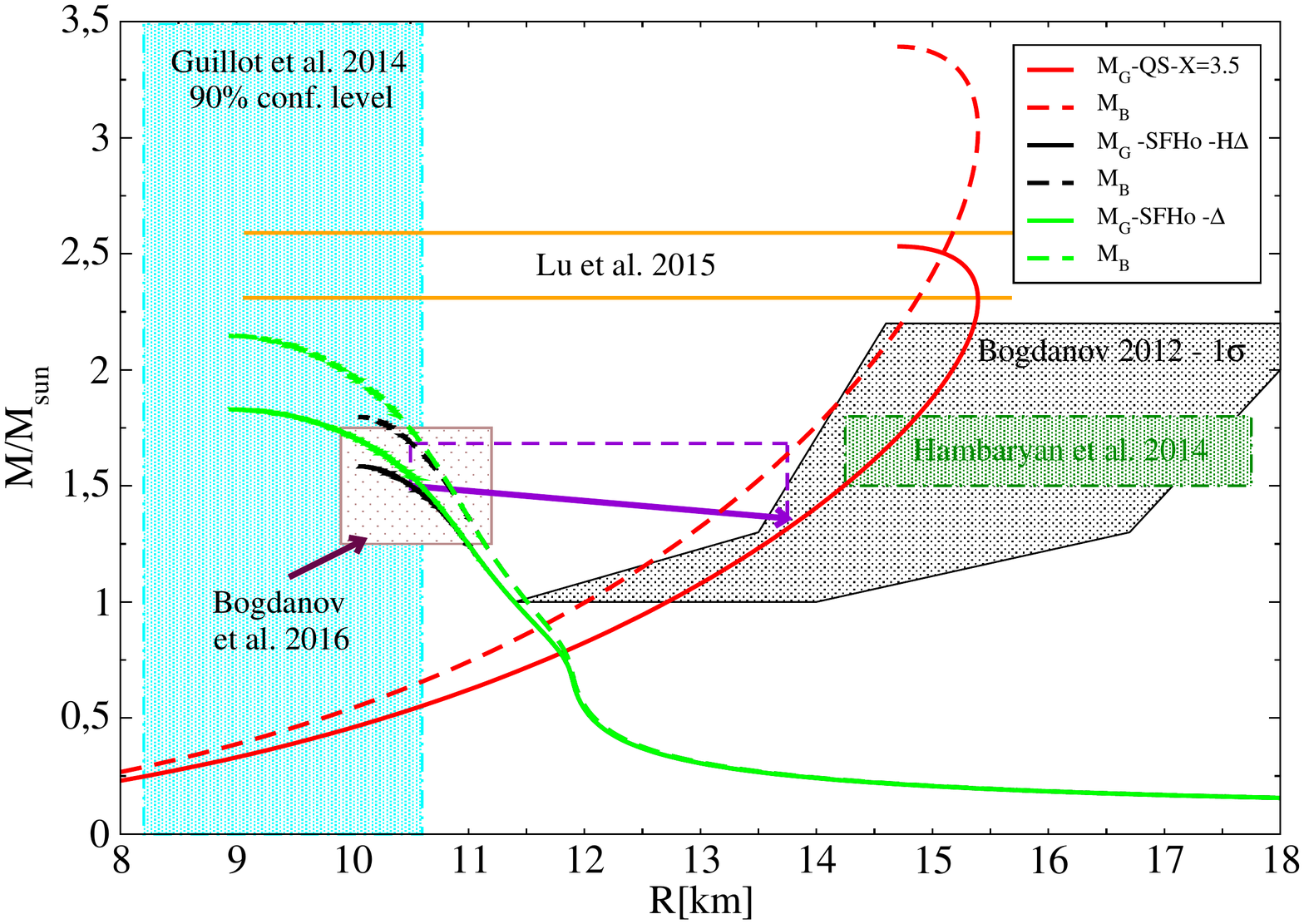}
    \caption{The mass-radius relations for hadronic stars and strange quark
        stars. The dashed lines represent the baryonic masses, whereas the solid
        lines represent the gravitational masses.  A typical transition for a NS
        reaching $M^H_{\rm max}$ is shown with violet color. The solid arrow
        starts at $M^H_{\rm max}$ and ends at $M^H_{\rm max}-\Delta M$. Although
        the baryonic masses before and after the transition are equal (dashed
        horizontal violet line), the QS gravitational mass is smaller ($M^H_{\rm
        max}-\Delta M = 1.36 M_{\odot}$) than the NS gravitational mass
        ($M^H_{\rm max}=1.5 M_{\odot}$). The maximum mass of the QS branch
        $M^Q_{\rm max}$ is significantly larger than $M^H_{\rm max}$.  The brown
        box is a rough approximation of the limits indicated in
        \citet{Bogdanov:2016nle}.}
    \label{fig:transition_Drago}
\end{figure}

To calculate the post-transition QS mass we implement the conservation of the
baryonic mass \citep{Bombaci:2000cv} while the gravitational mass changes due to
the different binding energies of NSs and of QSs (see
Fig.~\ref{fig:transition_Drago}).

The radius in the model we use is larger for a QS. It is irrelevant for the
present study, but quite crucial in the interpretation of the two-families
model. The maximum gravitational mass of a QS is not well-determined.  In our
calculations we assume the value of $2.5\msun$ (thus well above all known
massive NSs: \citet{Demorest1010,Antoniadis1304}). A possible way of
determining  $M_{\rm max}^Q$ is through the analyses of the extended emission
of short GRBs \citep{Lasky:2013yaa,Lu:2015rta,Li:2016khf}. In particular in
\cite{Lasky:2013yaa} the expected mass distribution for the post-merger remnant
is $M=2.46^{+0.13}_{-0.15} M_{\odot}$.  Although this limit includes also
supramassive stars, it represents a hint of the existence of stars with masses
significantly larger than $2M_{\odot}$.

A very crucial feature of our scheme is that the radius of the compact star
increases during the conversion. This is compatible with a combustion mode which
is not driven by pressure but by diffusion \citep{Olinto:1986je} and is strongly
accelerated by Rayleigh-Taylor hydrodynamical instabilities, as discussed in
\citet{Horvath:1988nb,Drago:2005yj,Herzog:2011sn,Pagliara:2013tza,Drago:2015fpa,Furusawa:2015vfn}.
However, these instabilities halt below a certain critical density and the
conversion of the most external layer is much slower, see also the recent
simulations of \citet{Ouyed:2017nuy}.

The present analysis does not contain two potentially relevant phenomena which
can take place in association with quark deconfinement.  First, the impact of
quarks deconfinement on the SN explosion is not discussed in this paper: we only
assume that if the compact star produced by the SN has a mass larger than
$M^H_{\rm max}$ then it immediately becomes a QS. On the other hand, quark
deconfinement could help heavy progenitors to explode
\citep{Drago:2008tb,Drago:2015dea}. This mechanism could in principle produce
compact stars with  higher masses.  Second, we do not take into account the
relation between mass accretion and angular momentum accretion. More explicitly,
in the present paper rotation is not considered.

\section{Results}

The results are scaled to be comparable with the Milky-Way equivalent galaxy
(MWEG), which we assumed to have a total stellar mass of $\mmweg
=\ttt{6.0}{10}\msun$ \citep[e.g.,][]{Licquia1506} and continuous star formation.
We chose $\mhmax=1.5\msun$, as our standard model and quantitative results refer
to this model unless differently stated. In Sec.~\ref{sec:mhmax} the effects of
changing the value of the maximum mass of hadronic stars to $\mhmax=1.6\msun$ are
analyzed. We show that the results and conclusions are qualitatively similar for
both models.

\begin{deluxetable}{lccccc}
    \tablecaption{Number of QS/NS in binaries}
    \tablehead{ Metallicity & \nqs\tablenotemark{a} &
    \nns\tablenotemark{a} & \fqs\tablenotemark{b} &
    \nns(noQS)\tablenotemark{c} & $\fcr$\tablenotemark{d} }
    \tablewidth{\columnwidth}
    \startdata
    \multicolumn{6}{c}{ALL}\\
    \zsun & \ttt{9.0}{4} & \ttt{7.2}{6} & $0.01$ & \ttt{7.3}{6} &
    $\mathbf{1.10}$ \\
    \zsun/10 & \ttt{2.7}{5} & \ttt{7.4}{6} & $0.04$ &\ttt{7.7}{6} &
    $\mathbf{1.37}$\\
    \zsun/100 & \ttt{1.5}{5} & \ttt{1.0}{7} & $0.01$ & \ttt{1.0}{7} &
    $\mathbf{1.57}$ \\
    \\
    \multicolumn{6}{c}{LMXB}\\
    \zsun & \ttt{1.6}{4} & \ttt{6.1}{4} & $0.26$ & \ttt{7.7}{4} & $\mathbf{1.61}$ \\
    \zsun/10 & \ttt{1.2}{4} & \ttt{1.5}{5} & $0.08$ &\ttt{1.6}{5} & $\mathbf{1.22}$\\
    \zsun/100 & \ttt{7.0}{3} & \ttt{2.1}{4} & $0.25$ & \ttt{2.9}{4} & $\mathbf{1.31}$\\
    \\
    \multicolumn{6}{c}{DQS/DNS}\\
    \zsun & \dash & \ttt{6.4}{5} & \dash & \ttt{6.6}{5} & $\mathbf{0.88}$ \\
    \zsun/10 & \ttt{4.2}{3} & \ttt{5.2}{5} & 0.08 & \ttt{5.2}{5} & $\mathbf{1.22}$\\
    \zsun/100 & \dash & \ttt{7.6}{5} & \dash & \ttt{7.6}{5} & $\mathbf{0.86}$\\
   
    \enddata
    \tablecomments{QS and NS quantities per MWEG at present time for
        $\mhmax=1.5\msun$. ALL\dash all binaries; LMXB\dash mass-transferring
        binaries; DQS/DNS\dash double QS/NS.}
    \tablenotetext{a}{Number of QS (\nqs) and NS (\nns)}
    \tablenotetext{b}{fraction of QSs; defined as $\fqs:=\nqs/(\nqs+\nns)$}
    \tablenotetext{c}{number of NSs in the model without QSs (noQS)}
    \tablenotetext{d}{change in a number of compact objects (QSs and NSs)  in
        $1.36\dash1.5\msun$ mass range; $\fcr:=(\nqs'+\nns')/\nns'(noQS)$ (mass
        range marked with $'$)} \label{tab:results}
\end{deluxetable}

The ratio of the number of QSs to NSs is between $0.01\dash0.04$ depending on
metallicity (Tab.~\ref{tab:results}), but for mass-transferring binaries (in the
case of LMXBs) it is higher: $0.08\dash0.26$. This corresponds to
$0.9\dash\ttt{2.7}{5}$ QSs in a MWEG. Most of them are existing in wide and
therefore non-interacting binaries, or are in pairs with low-luminosity
companions, which in most cases ($78\dash95\%$) are white dwarfs (WD).

\begin{figure}[h]
    \centering
    \includegraphics[width=\columnwidth]{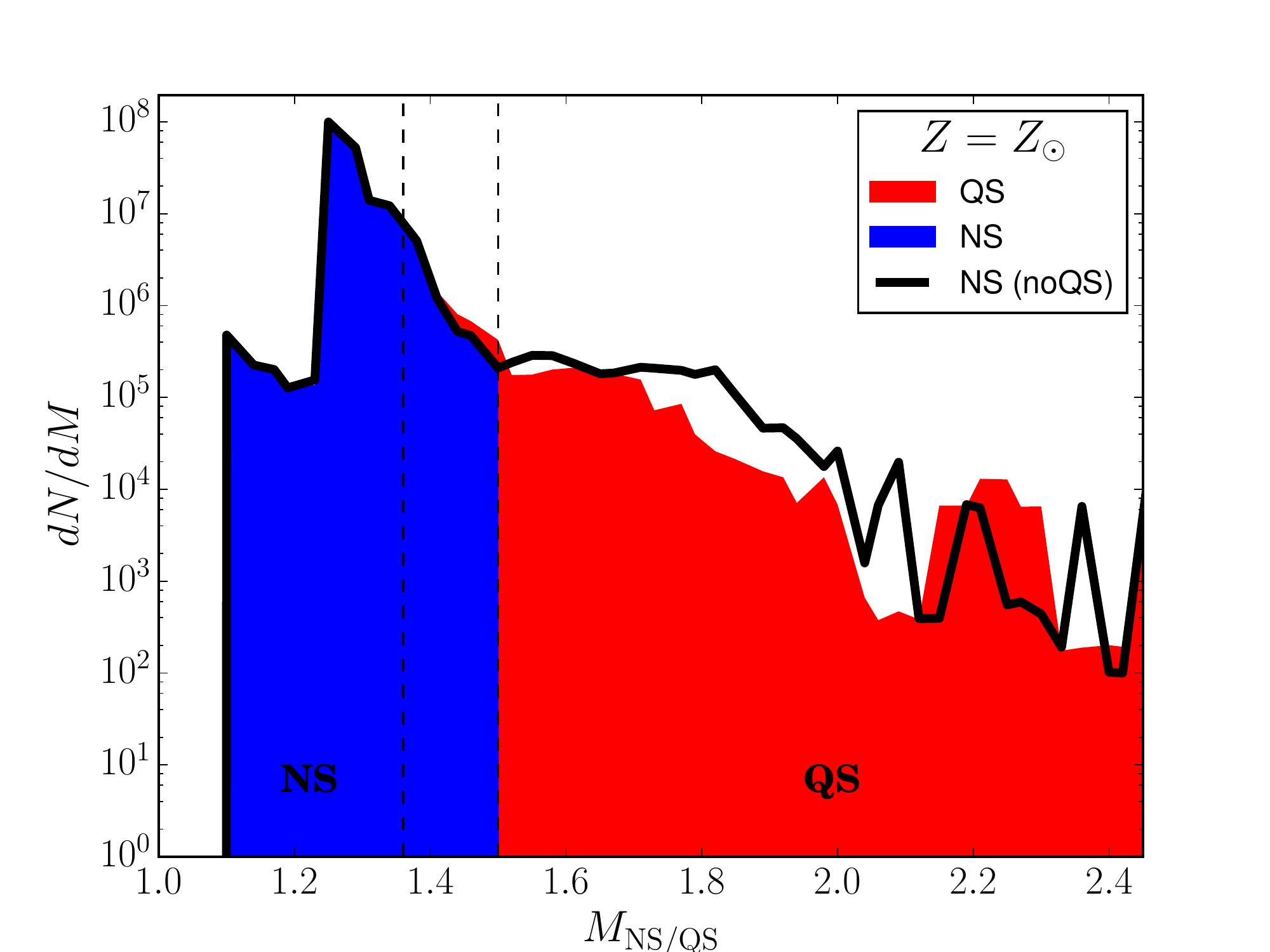}\\
    \includegraphics[width=\columnwidth]{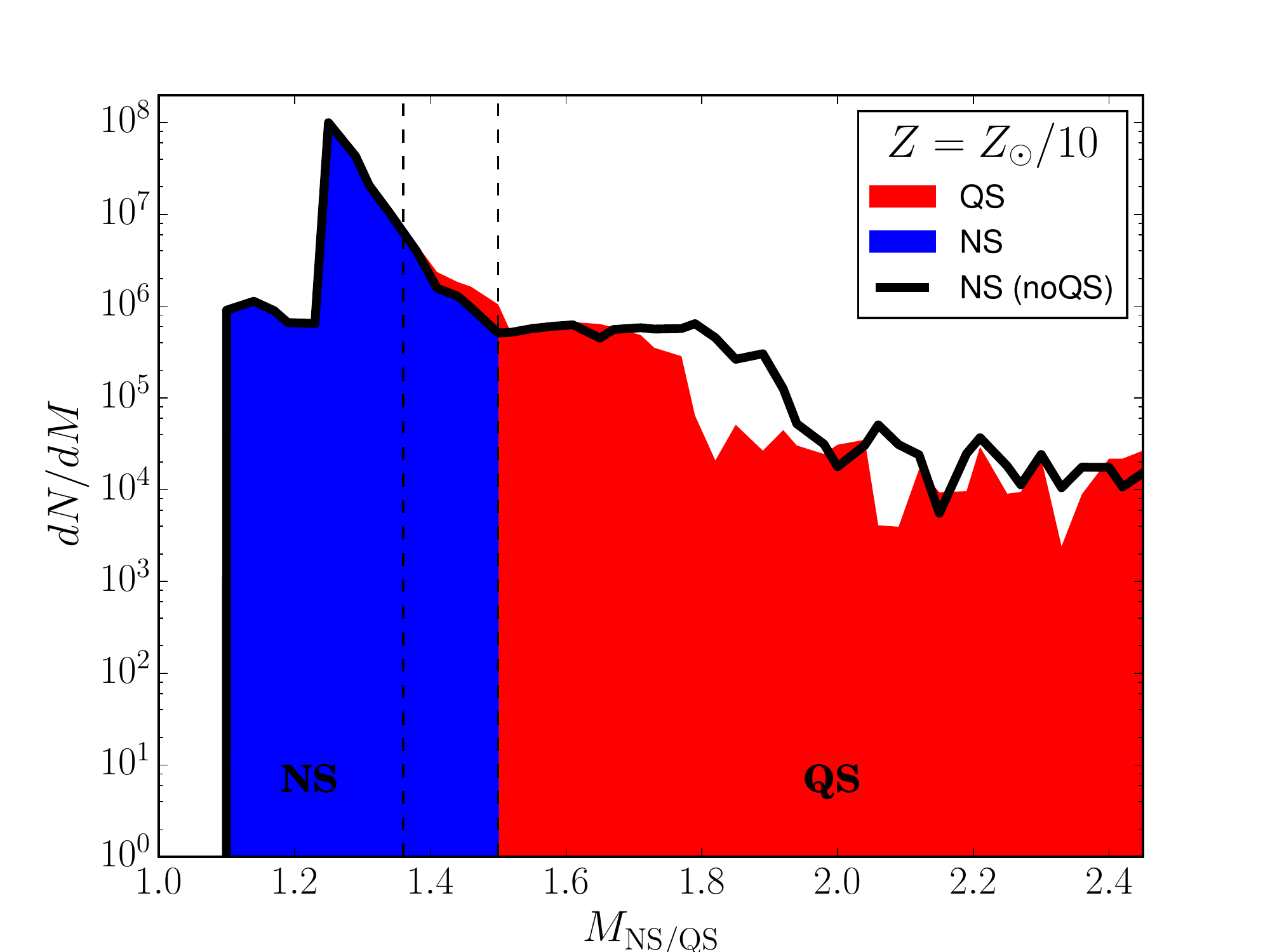}\\
    \includegraphics[width=\columnwidth]{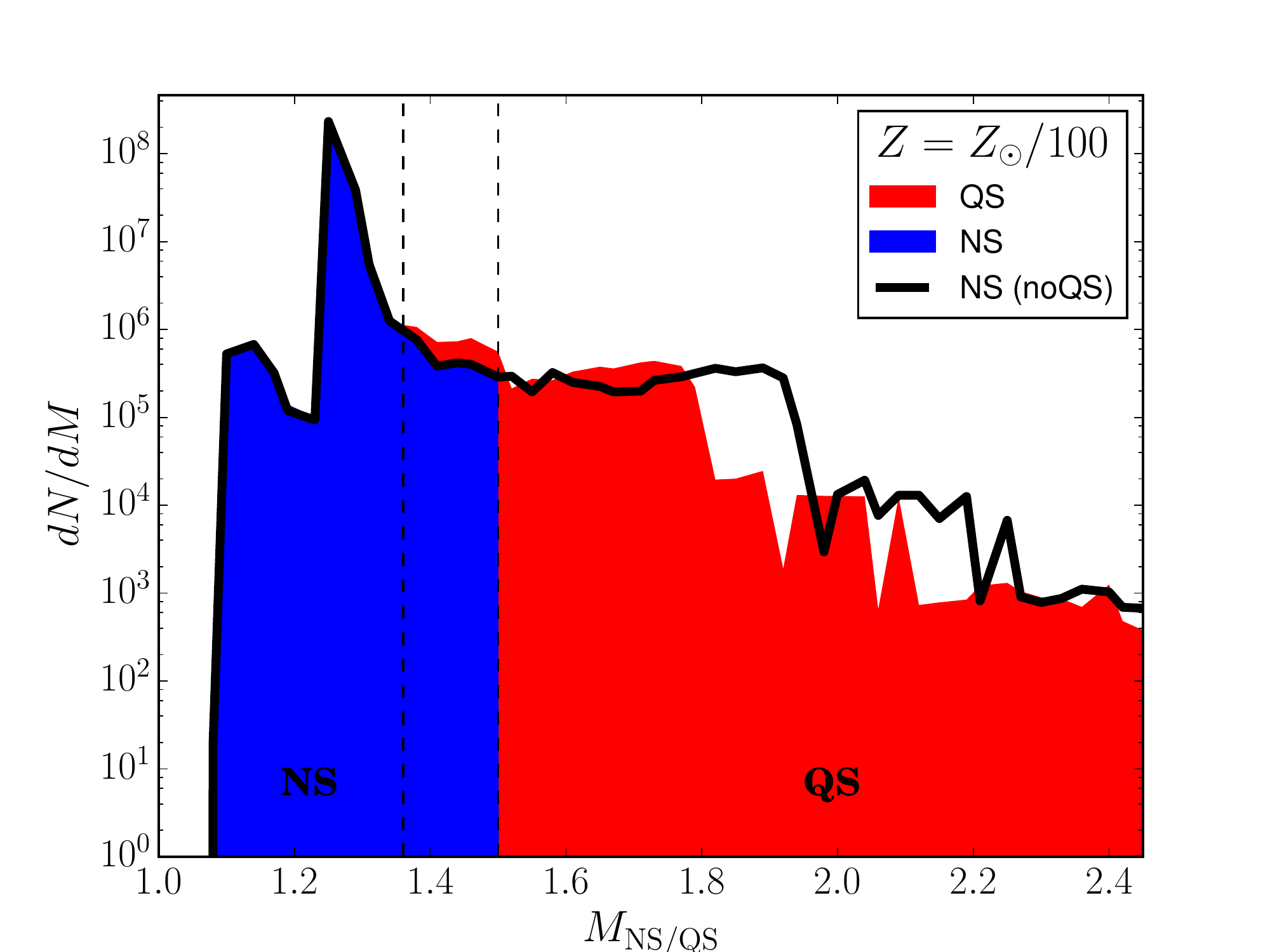}
    \caption{The distribution of masses for NS (blue) and QS (red) for
        metallicity $Z=\zsun$ (upper plot), $Z=\zsun/10$ (middle plot), and
        $Z=\zsun/100$ (lower plot). The black line marks the distribution of NSs in
        model without QSs (noQS). Features seen in the QS mass distribution are related
        to post-QS-formation evolution and are not a subject of this study.}
    \label{fig:qsmasses_all}
\end{figure}

Fig.~\ref{fig:qsmasses_all} demonstrates the effect of deconfinement on the
compact stars' (QS/NS) mass distribution in comparison to the model without
deconfinement (noQS).  Notice specifically that in the range $1.1\dash1.5\msun$,
where most of the NSs reside, the difference is minimal.  In particular, the
deconfinement does not affect the peak in the distribution at $\sim1.3\msun$.

\subsection{Formation of QSs}\label{sec:formation}

\begin{deluxetable*}{llccc}
    \tablecaption{Formation of strange quark stars in binaries}
    \tablehead{ &  Typical evolutionary route\tablenotemark{a} & \multicolumn{3}{c}{\#QS\ per MWEG}  \\ & & \zsun & \zsun/10 & \zsun/100 }
    \tablewidth{\textwidth}
    \startdata
    \raacc & CE1(6-1;12-1) MT2(12-3) AICNS1 MT2(13-3) AICQS1 & \ttt{8.6}{4} & \ttt{2.1}{5}& \ttt{1.2}{5} \\
    \radir & CE1(4/5-1;7/8-1) SNQS1 & \ttt{4.0}{3}& \ttt{2.6}{4}& \ttt{3.2}{3} \\
    \rbdir & MT1(2-1) SN1 CE2(14-4;14-7) SNQS2 & \ttt{6.5}{2}& \ttt{2.8}{4}& \ttt{2.7}{4} \\
    \\
    \rlmxb & CE1(6-1;12-1) CE2(12-3;12-7) MT2(12-7) AICNS1 MT2(13-7/8) AICQS1 MT2(13-11/17) & \ttt{1.6}{4} & \ttt{1.2}{4} & \ttt{7.0}{3} \\
    \\
    \rdqs  & MT1(4-4) CE2(7-4;7-7) SNQS1 SNQS2 & \dash & \ttt{4.2}{3} & \dash
    \enddata
    \tablecomments{QS formation channels: \raacc\dash QS forms from a NS due to
        mass-accretion; \radir\dash QS form directly after SN; \rbdir\dash QS forms
        from a secondary (less-massive star on ZAMS); \rlmxb\dash QS in LMXB
        (mass-transfer present); \rdqs\dash double QS.}
    \tablenotetext{a}{Only most important evolutionary phases are present:
        MT1/2~--~mass transfer from the primary/secondary; CE1/2~--~common
        envelope (primary/secondary is a donor); AICSN1 -- accretion induced
        collapse of a WD into a NS; AICQS1~--~accretion induced collapse of a NS
        into a QS; SNQS1/2~--~ direct formation of a QS after supernova of the
        primary/secondary. Stellar types: 1 -- main sequence; 2~--~Hertzsprung
        Gap; 3 -- red giant; 4 -- core He-burning; 5/6 -- early/thermal pulsing
        asymptotic giant branch; 7 -- He star; 8 -- evolved He star; 11~--~
        Carbon-Oxygen White Dwarf; 12~--~Oxygen-Neon white dwarf; 13~--~neutron star;
        14~--~black hole; 17~--~Hybrid white dwarf.}
    \label{tab:formation}
\end{deluxetable*}

\begin{deluxetable*}{cccccc}
    \tablewidth{\textwidth}
    \tablecaption{Typical parameters for formation channels}
    \tablehead{Parameter\tablenotemark{a} & \raacc & \radir & \rbdir & \rlmxb & \rdqs\tablenotemark{b}}
    \startdata
    $\ma\,[\msun]$ 	& $1.4\dash1.8$		& $1.4\dash1.7$		& $1.4\dash1.8$		& $1.4\dash1.6$	& $1.4\dash1.6$\\
    $\mb\,[\msun]$ 	& $0.3\dash0.4$		& $\lesssim0.7$ 	& $7.8\dash30$		& $\lesssim0.2$	& $\sim1.7$\\
    $a\,[\rsun]$ 	& $\lesssim170$		& $\lesssim4600$ 	& $\lesssim540$		& $\lesssim1.9$	& $7.5\dash24$\\
    $\tage\,[\myr]$ 	& $\gtrsim8000$		& $4.2\dash6000$ 	& $5.1\dash6500$	& $900\dash2600$	& $140\dash6600$\\
    \\
    $\mzamsa\,[\msun]$ & $6.1\dash7.8$		& $16\dash28$		& $18\dash30$			& $6.0\dash12$	& $21\dash24$\\
    $\mzamsb\,[\msun]$ & $1.0\dash1.5$		& $2.0\dash4.1$		& $41\dash77$			& $0.7\dash4.7$	& $20\dash23$\\
    $\azams\,[\rsun]$ 	& $560\dash2200$	& $2900\dash4500$	& $560\dash8000$		& $700\dash2700$	& $570\dash1200$\\
    \enddata
    \tablecomments{The table presents the typical values of strange quark star
        and companion masses and their separation for the present time and ZAMS.
        In case of the present time, the age of the system is also provided.
        Scenarios' designations are explained in Tab.~\ref{tab:formation} and in
        Sec.~\ref{sec:formation}. }
    \tablenotetext{a}{\ma~--~QS mass; \mb~--~companion mass; $a$~--~separation;
        $\tage$~--~age of the system (time since ZAMS)} 
    \tablenotetext{b}{Both components are QSs}
    \label{tab:params}
\end{deluxetable*}

Mostly, a QS forms from a primary (i.e. the more massive component in the binary
on ZAMS). In $\sim72\dash96\%$ of cases a QS forms as a consequence of MT onto a
NS (route \raacc). Only the remaining $\sim2\dash10\%$ of QSs form directly
after a SN explosion (\radir). Rarely ($\leq20\%$), a QS may form from a
secondary, if it is initially massive (\rbdir; $\sim18\dash30\msun$). Most of
these QSs do not interact with their companions, but in about $3\dash18\%$ of
cases a LMXB can form (\rlmxb; Sec.~\ref{sec:lmxb}).  If the metallicity is
moderate ($Z=\zsun/10$) double QSs form as a result of binary evolution (\rdqs;
Sec.~\ref{sec:dqs}). Tables \ref{tab:formation} and \ref{tab:params} summarize the most typical
evolutionary routes for all scenarios. Although models for different
metallicities share the same trends, there are differences in the total number
and relative abundances of QSs formed via different channels.

\paragraph{\raacc; QS forms through accretion onto a NS} This is the most
typical formation scenario of QSs in binaries. In a typical case, a primary is
about $7.2\msun$ and a secondary is $1.2\msun$. The primary evolves faster and
fills the Roche lobe (RL) while being on the asymptotic giant branch after
$53\myr$. The MT is usually unstable due to large mass ratio and a common
envelope (CE) occurs. If the binary survives this phase, the primary is ripped
off its hydrogen envelope and becomes an Oxygen-Neon WD with a mass of about
$1.3\msun$.  Afterwards, the secondary evolves, becomes a red giant (RG) and
fills its RL. A MT from the secondary increases the mass of the WD up to
$1.38\msun$. Then the primary collapses and becomes a $1.26\msun$ NS.
Afterwards, the secondary re-fills the RL and commences a MT again. The system
becomes a LMXB. The NS mass may rise up to $1.5\msun$ due to accretion and the
deconfinement transforms it into a QS with a mass of $1.36\msun$ (Fig.
\ref{fig:evroutes}, upper plot).  The MT may proceed further, what will allow
the QS to reach a higher mass (typically up to $1.8\msun$). The evolution leads
to the formation of a QS-WD system, which is usually too wide to interact
anymore.

As already remarked before, here we do not consider the effect of rotation on
the structure of compact stars.  As discussed in \citet{Bejger:2011bq}, the
central density during mass accretion could increase marginally due to the
simultaneous increase of the angular momentum.  Therefore the conversion of the
NS could occur either during the mass accretion stage or after the end of mass
accretion during the spin down.

\paragraph{\radir; Direct collapse to a QS after a SN explosion} The initial
binary is more massive than in \raacc\ scenario. The primary's initial mass is
about $16\dash28\msun$ and the secondary's mass is $\sim2.0\dash4.1\msun$. When
the primary fills its RL, the MT is unstable, so the CE phase commences. The
secondary is massive enough to eject the envelope and the system survives with a
much shorter orbit (due to orbital angular momentum loss). Additionally, the
outer envelope of the primary is ripped off. The SN explosion, which occurs
shortly after, may significantly change the separation. Usually no further
interaction is observed and the secondary evolves unaffected and forms a WD
after $\sim1\gyr$. 

\bigskip

The main factor that affects the ratio of QS formation in two above routes is
the mass of the components on ZAMS. Progenitors of QSs in \raacc\ are lighter,
therefore, are more abundant (approximately twice) in the initial populations,
then in the case of \radir. Additionally, in the case of a heavy primary
(\radir) it is hard for the secondary to eject the massive envelope of a primary
during CE phase, which frequently leads to a merger. On the other hand, without
a CE a system is still wide during a SN. Consequently, it is frequently
disrupted (in $\sim98\%$ of cases).

The companions in \raacc\ are usually low-mass WDs, in agreement with the
results by \citet{Popov0505}, who found that many of the most massive
uncollapsed compact objects might be observed at the stage of accretion from WD
companions.  The QS is formed at the stage of accretion which lasts for a long
time. After accretion is over the QS is spun-up and can be observed as a
millisecond radio pulsar.

\paragraph{\rbdir; QSs from secondaries} QSs formed from secondaries, i.e. less
massive binary components on ZAMS, constitute only a small fraction ($\leq20\%$)
of the QS population and the fraction is in general larger for lower metallicity
(Tabs.~\ref{tab:formation} and \ref{tab:comparison}). Primaries in these systems
evolve faster than the QS progenitors and usually become BHs.

On ZAMS the binary consist of a $\sim41\dash77\msun$ primary and
$\sim18\dash30\msun$ secondary. The separation is short enough that when the
primary enters the Hertzsprung-Gap (HG) phase its RL is filled. The  MT is
non-conservative. A BH forms directly after a SN explosion with small mass loss
(typically, BHs obtain low natal kicks). Its mass is between $7.8\dash30\msun$.
A few \myr\ later, the secondary fills its RL while being on the core helium
burning. The CE occurs in which the separation is shortened and the star
losses its outer hydrogen envelope. The second SN results in a direct formation
of a QS and the binary becomes a double compact object (BH-QS).

\subsection{Low-mass X-ray binaries}\label{sec:lmxb}

We performed a separate analysis for accreting QS. We considered all
mass-transferring binaries with NS/QS accretors. Initial donor masses are in
general below $3\msun$, as heavier companions usually provide dynamically
unstable MT. Properties of X-ray emission from accreting NSs or QSs are similar
\citep{Alcock8611}, therefore, accreting QSs constitute a subgroup of LMXBs.

Although most of the QSs form through accretion from a RG companion (\raacc),
the MT phase is relatively short and the resulting QS-WD binary is too wide to
commence Roche lobe overflow. Consequently, the evolutionary route leading to
the formation of a LMXB with a QS is different. We found that a typical
companion is a WD with a very small mass ($\lesssim0.2\msun$). It is a direct
consequence of a prolonged mass transfer onto the primary. The separation which
allows for a Roche lobe overflow is small ($\lesssim1.9\rsun$) and the period is
very short ($P\lesssim6$ h).

\paragraph{\rlmxb; Accretion onto a QS} The initial evolution towards the
formation of a LMXB with a QS is in general similar to \raacc, however,
secondaries are on the average more massive on ZAMS. As a result, the heavier RG
companion after the first CE provides unstable MT, which results in a second CE
phase (this time the secondary is a donor). The orbit shrinks and the secondary
loses much of its mass ($\sim80\%$). The resulting helium star
($\mb\approx0.7\msun$) re-fills the RL and transfers mass onto a heavy WD
primary ($\ma\approx1.3\msun$). Afterwards, the latter becomes a NS and after
another $20\myr$ a QS. As a result of the mass loss, the secondary becomes a WD.
The separation is very small ($\sim0.3\rsun$) due to two earlier CE phases, so
the secondary is able to fill its RL again due to gravitational radiation (GR).
A long and stable MT phase proceeds during which the mass of the donor drops
below $0.2\msun$ (Fig.~\ref{fig:evroutes}, middle plot).

\begin{figure}[h]
    \centering
    \includegraphics[width=\columnwidth]{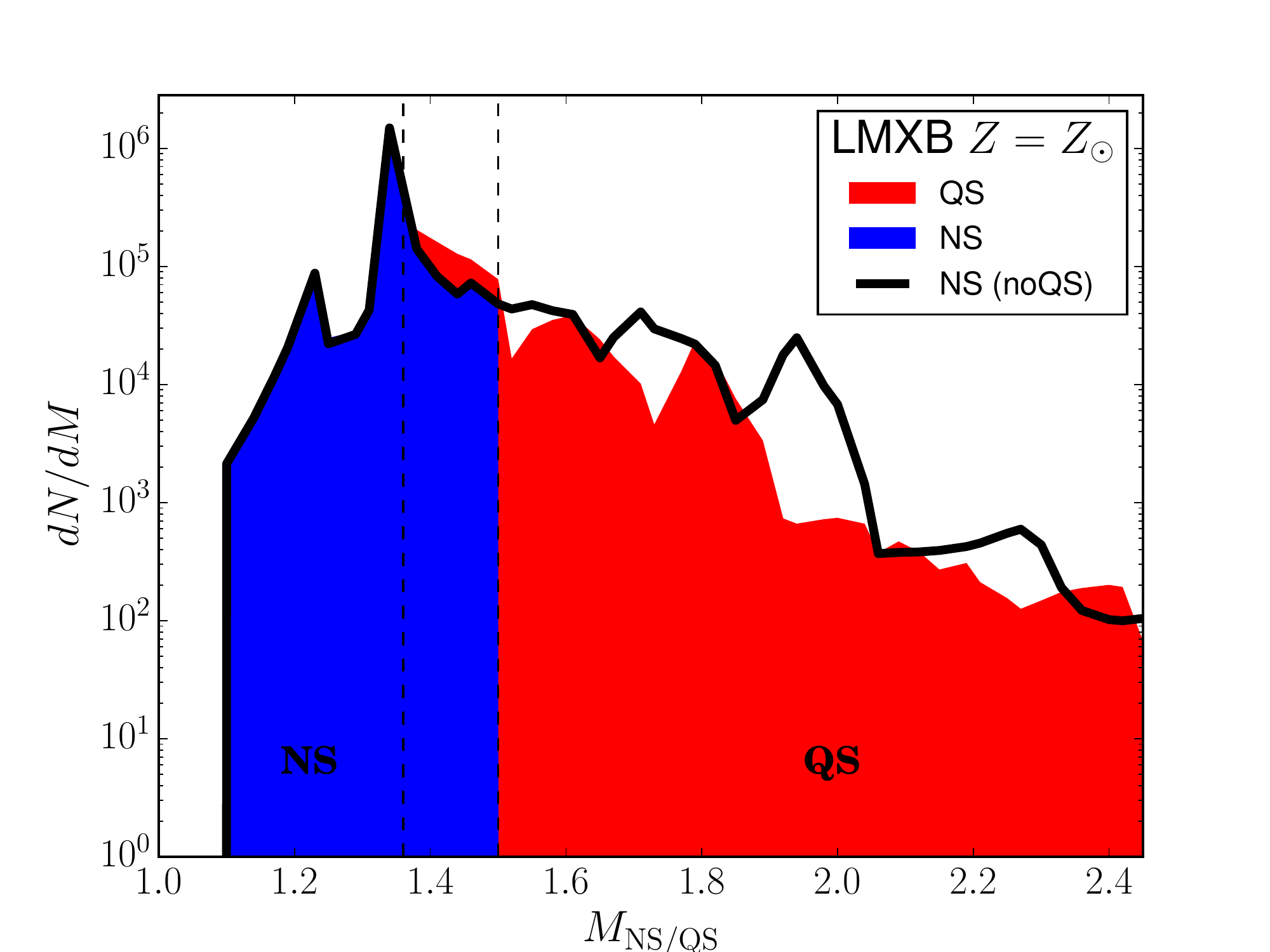}\\
    \includegraphics[width=\columnwidth]{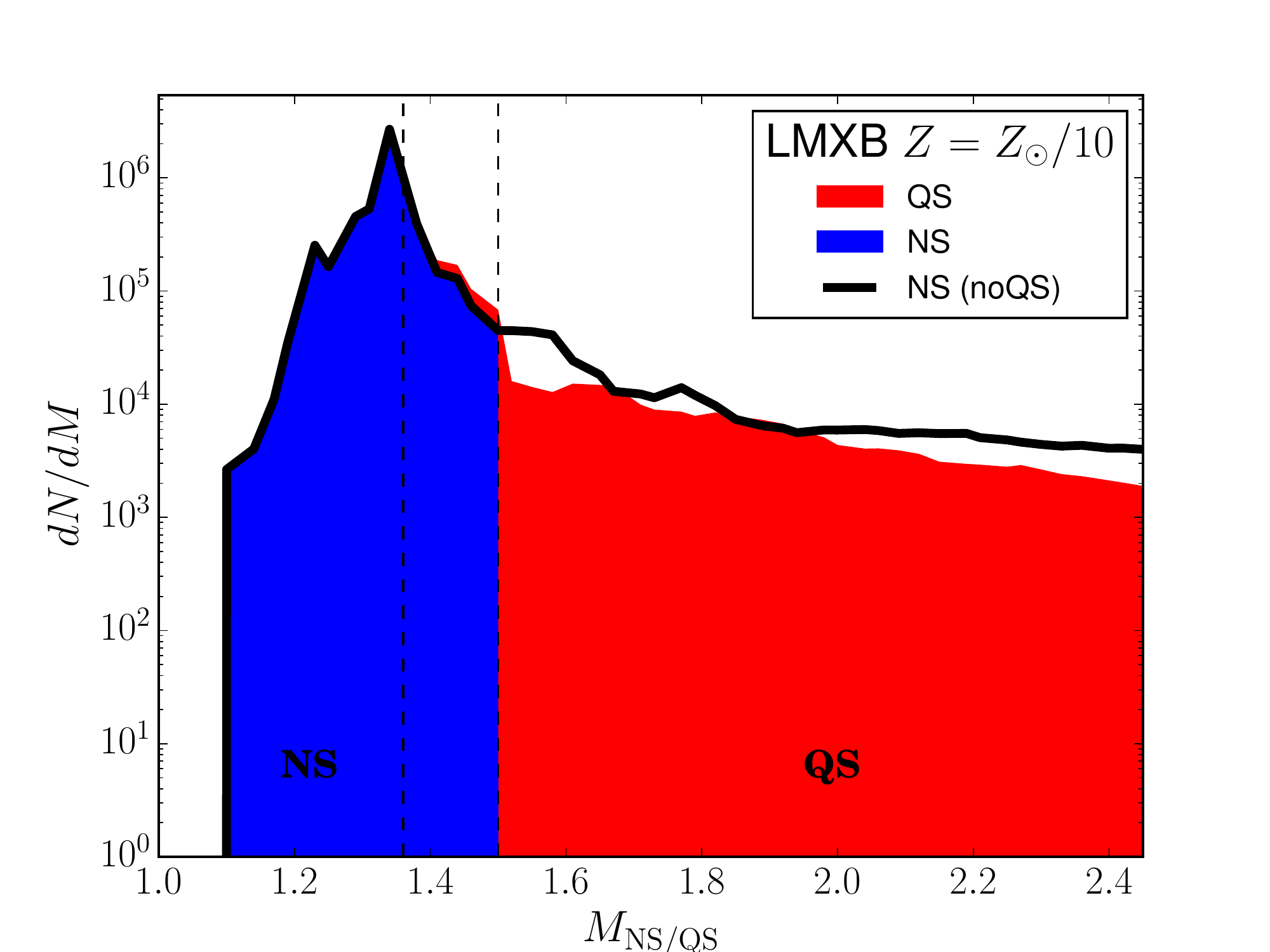}\\
    \includegraphics[width=\columnwidth]{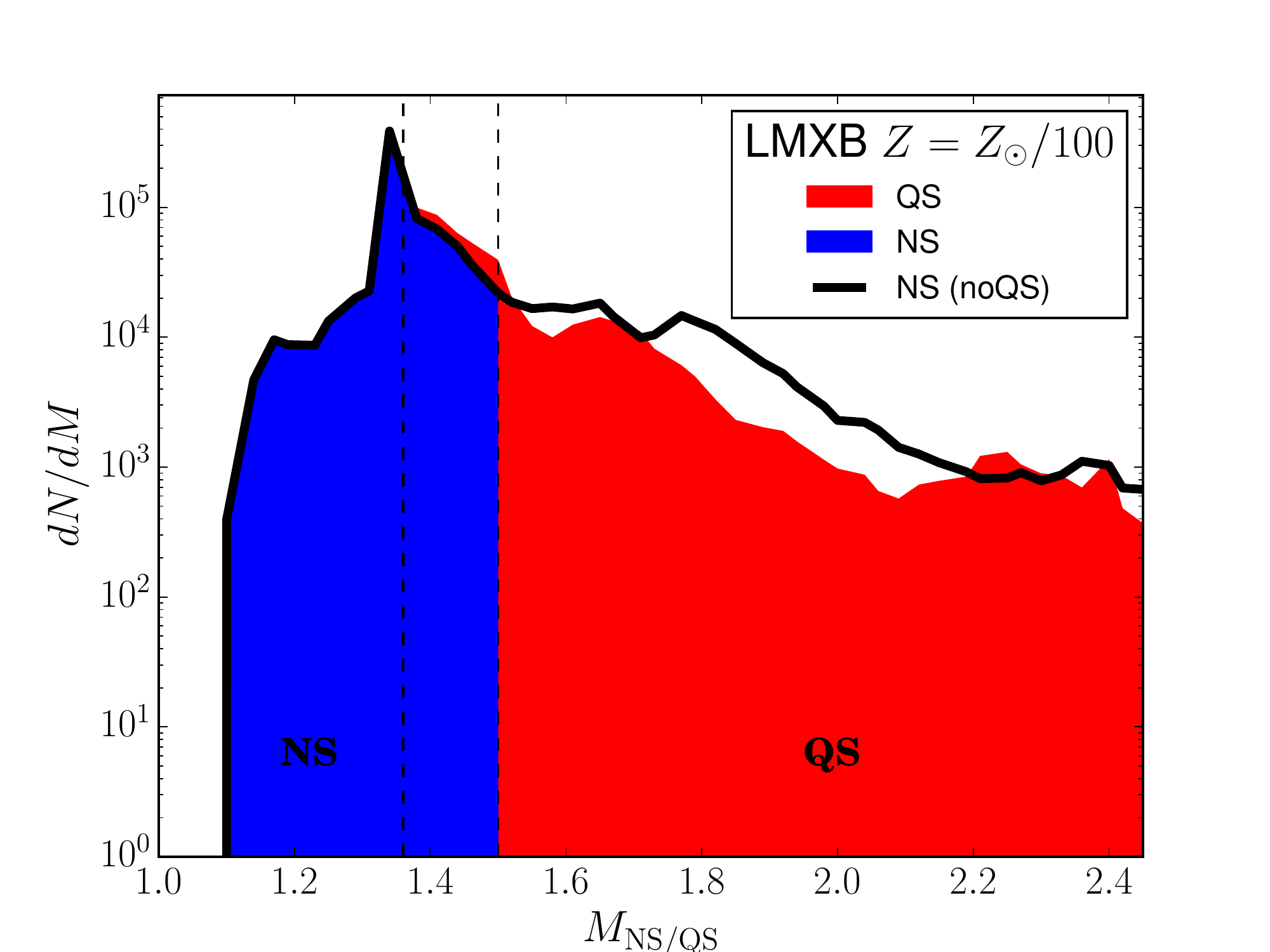}
    \caption{The distribution of masses for NS (blue) and QS (red) in LMXBs for
        metallicity $Z=\zsun$ (upper plot), $Z=\zsun/10$ (middle plot), and
        $Z=\zsun/100$ (lower plot). The LMXB is defined as a mass-transferring
        binary with NS/QS accretor. The black line marks the distribution of
        NSs' masses in model without QSs (noQS). Features seen in the QS mass
        distribution are related to post-QS-formation evolution and are not a subject of
        this study.}
    \label{fig:qsmasses_lmxb}
\end{figure}

The mass distributions of compact objects in LMXBs
(Fig.~\ref{fig:qsmasses_lmxb}) differ from those including all binary systems
(Fig.~\ref{fig:qsmasses_all}).  In models involving formation of QSs, we
obtained a higher number of systems ($22\dash67\%$) in the coexistence range. In
general, the increase is more significant for higher values of \mhmax. In spite
of this excess, the main peak of the mass distribution is still at
$\sim1.3\msun$ (so outside of the coexistence range) and its position and
magnitude are unaffected by the deconfinement. Indeed, most of the mass
measurements are outside of the coexistence range, thus we cannot investigate
the presence of this difference.

\subsection{Double QSs}\label{sec:dqs}

The main hindrance to the formation of double compact objects is the natal kick
that may disrupt the binary during either of SN explosions. Nevertheless, we
found an evolutionary route leading to the formation of double QS (DQS).
Noteworthy, such scenario may be realized only in  stellar populations with
moderate metallicity ($Z=\zsun/10$). Only in such an environment we will observe
a higher number of DQS/double NS in the coexistence range in comparison to model
without deconfinement ($\fcr$; see Tab.~\ref{tab:results}). For \zsun, or
\zsun/100 metallicity, the $\fcr$ fraction is $<1$, which marks the fact that
the deconfinement in general hinders the formation of double compact objects by
$\sim0.1\dash3.0\%$.

\begin{figure}
	\includegraphics[width=0.9\columnwidth]{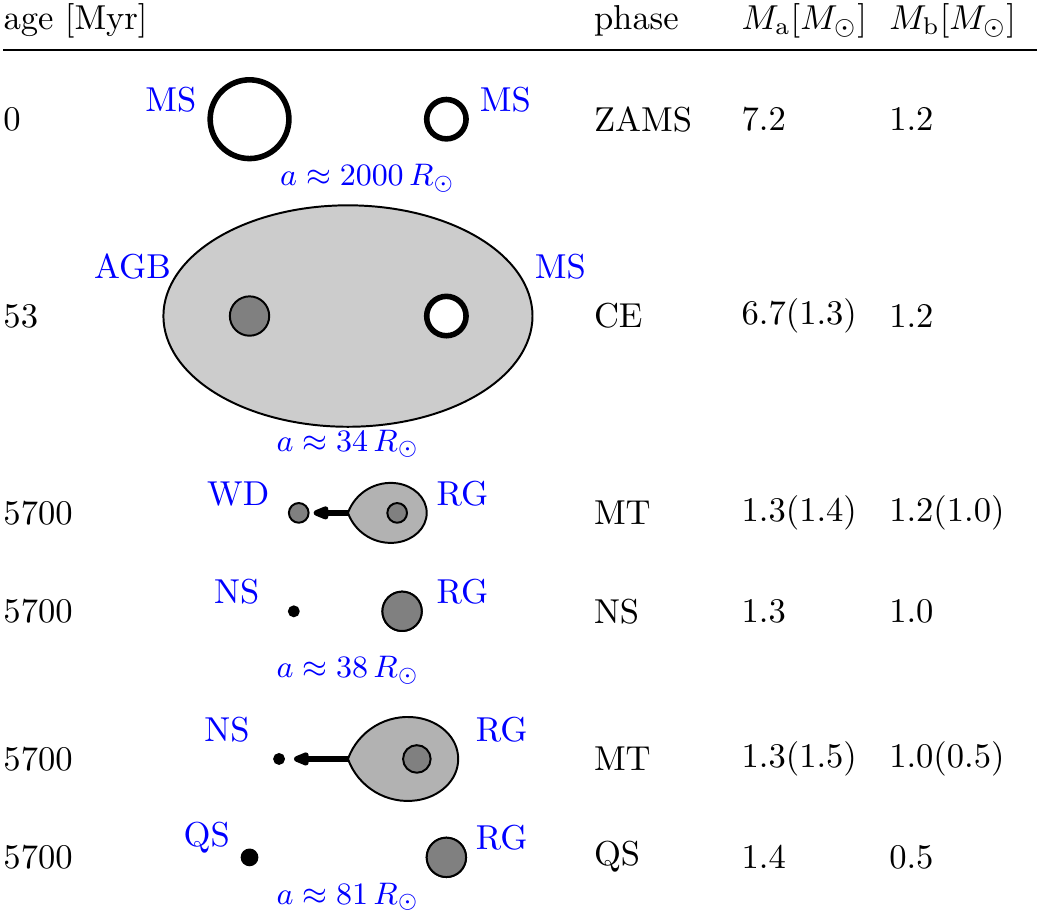}
    \includegraphics[width=0.9\columnwidth]{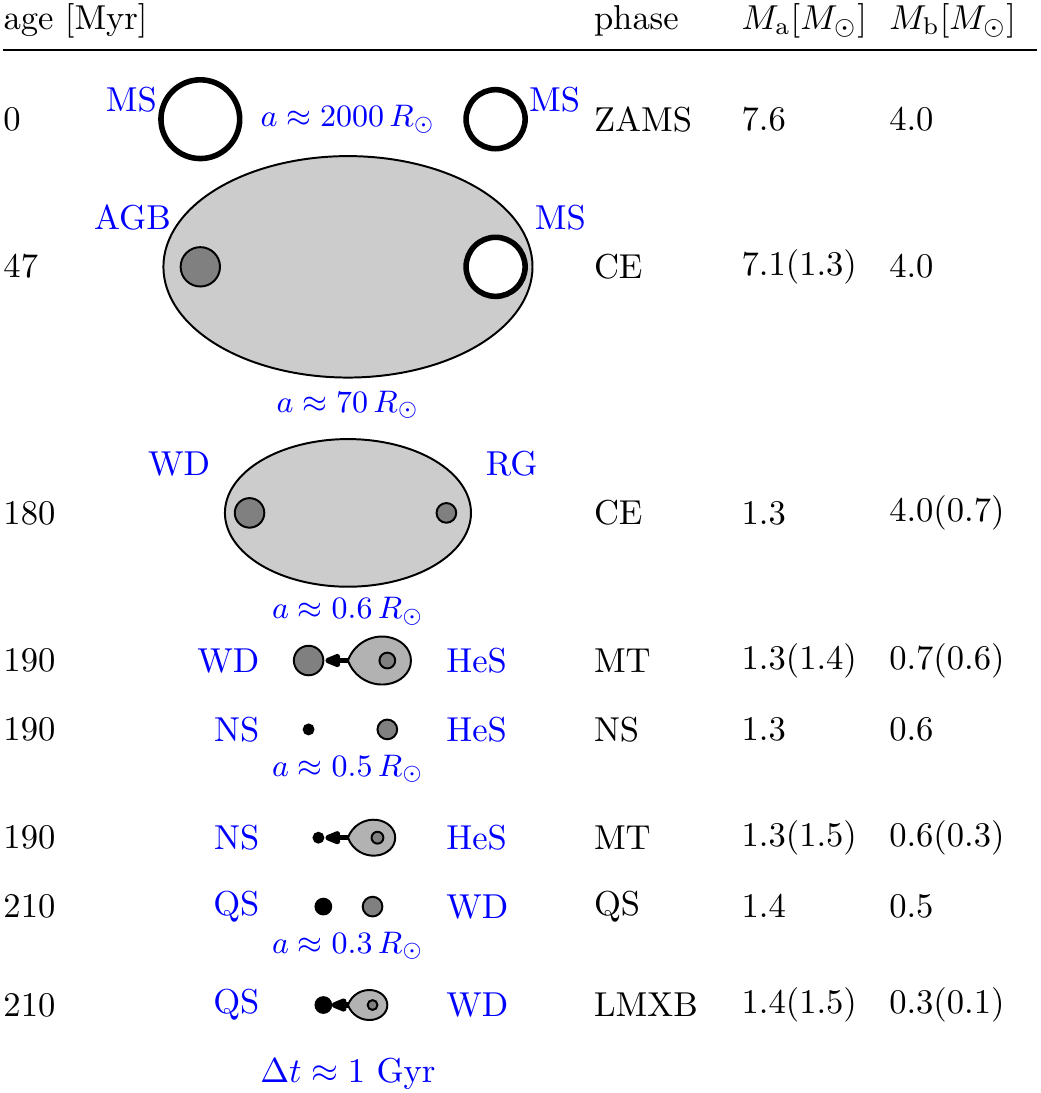}
    \includegraphics[width=0.9\columnwidth]{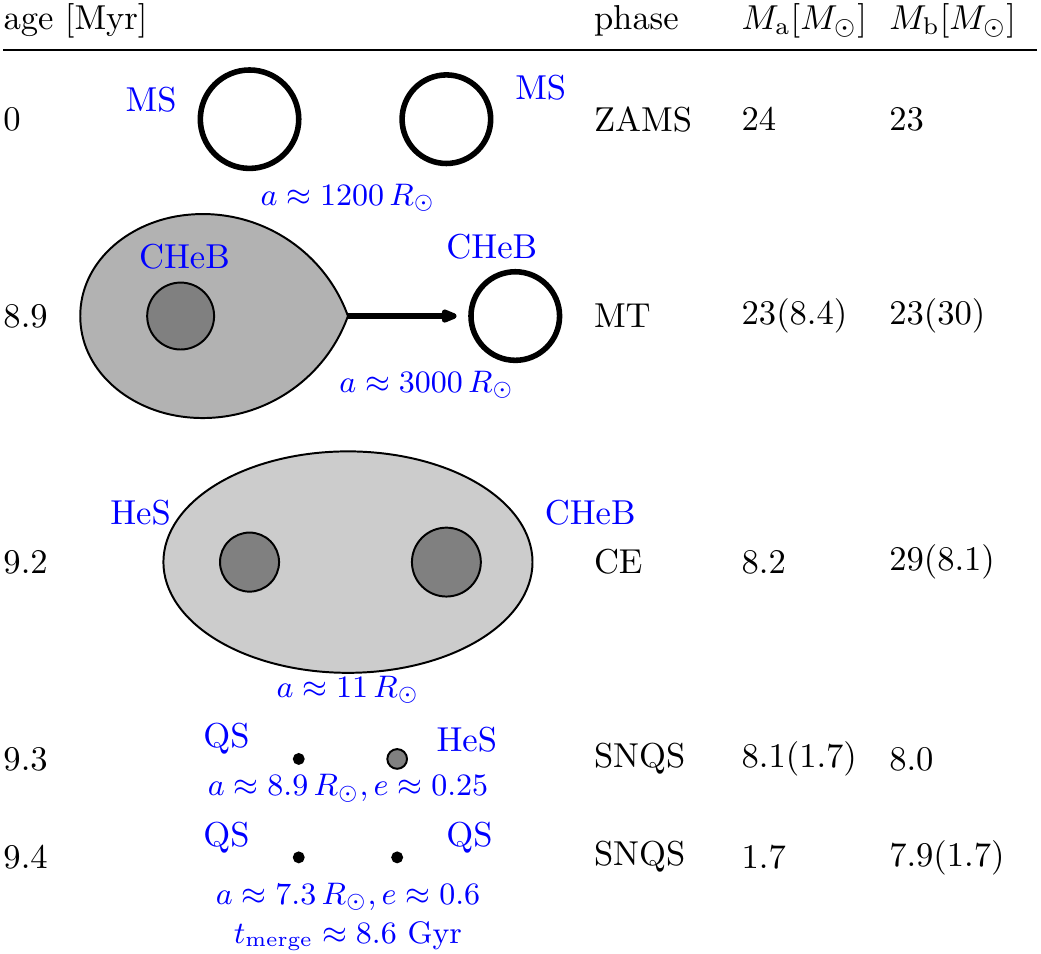}    
    \caption{Schematic representation of a typical binary evolution leading to
        the formation of a QS (upper plot), LMXB with a QS (middle plot), or a
        double QS (lower plot). $a$ is a separation and $e$ is an eccentricity.
        Abbreviation are explained in Tab.~\ref{tab:formation}. For details see
        Sec.~\ref{sec:formation}.}
    \label{fig:evroutes}
\end{figure}

\paragraph{\rdqs; Double QS binary} Typically, a DQS originates from a binary
which on ZAMS consists of two stars with masses $\sim20\dash24\msun$, where the
primary is on the average only slightly ($\sim1\msun$) heavier than the
secondary. The orbit must be wide enough to accommodate these stars
($570\dash1000\rsun$). In a typical system, at the age of $\sim9\myr$, the
primary fills the RL during the core helium burning phase and transfers mass
onto the secondary. After the MT phase, the primary becomes an $\sim8.2\msun$
helium star with $\sim29\msun$ core helium burning companion. After $200\kyr$,
the secondary expands, fills the RL and commences the CE phase. The separation
shrinks to a few \rsun\ and the double helium star forms. The primary and the
secondary sequentially (after 300 and 500 \kyr, correspondingly) explode and
form two QSs directly (like in \radir\ scenario). The system have a high chance
of surviving and forming a DQS on an orbit of $7.5\dash24\rsun$
(Fig.~\ref{fig:evroutes}, lower plot).

The presented scenario does not work for solar metallicity (\zsun). The reason
for that is a strong expansion of high-$Z$ stars. The RL is filled earlier for
the same initial separation (when the primary is on the HG), then for
$Z=\zsun/10$ (primary is on a core helium burning). Consequently, the mass-loss
is faster and longer, so the helium star forms earlier. This results in the SN
explosion happening before the CE phase may commence and this shortens
significantly the separation. Therefore, the orbit is larger ($\sim1000\rsun$)
in comparison to post CE systems ($\sim10\rsun$), what leads to a system
disruption during SNe.

There are no DQS in the lowest-$Z$ environments ($\zsun/100$) neither, but for a
different reason. Low-metallicity stars have small wind mass-loss rate.
Additionally, their expansion rate is smaller than for higher-$Z$ stars. This
results in a lower chance of MT occurrence during the core helium burning phase.
As a consequence, after the MT phase, the primary is still massive and forms a
BH. For lower-mass primaries the problem is still present as the expansion is
small and it is hard to commence a CE phase and to shrink the orbit before the
second SN. As a result, the system becomes disrupted when the second SN occurs. 

The estimated merger rate of DQSs due to GR \citep{Peters6411} is $\sim71$
events $\gyr^{-1} \mmweg^{-1}$. An average time to coalescence after the
formation of the second QS is about $10,000\gyr$ and only a few systems are
compact enough to merge within the Hubble time. In the case of the MW galaxy,
low-$Z$ stars are present mainly in the  bulge, which constitutes only $\sim1/6$
of the Galactic mass \citep{Licquia1506}. Therefore, we estimate the merger rate
for the MW as $\sim12$ events $\gyr^{-1}$ assuming a constant star formation
rate.

\subsection{Importance of the value of the limiting gravitational mass of NSs (\mhmax)}\label{sec:mhmax}

\begin{deluxetable*}{lcccccc}
    \tablewidth{\textwidth}
    \tablecaption{Comparison of models with $\mhmax=1.5\msun$ and $1.6\msun$}
    \tablehead{ & \multicolumn{2}{c}{$Z=\zsun$} & \multicolumn{2}{c}{$Z=\zsun/10$} & \multicolumn{2}{c}{$Z=\zsun/100$} \\ $\mhmax[\msun]$ & 1.5 & 1.6 & 1.5 & 1.6 & 1.5 & 1.6}
    \startdata
    \nqs   & \ttt{9.0}{4} \hfill $(1\%)$ & \ttt{4.6}{4} \hfill $(1\%)$ &    \ttt{2.7}{5} \hfill $(4\%)$ & \ttt{1.3}{5} \hfill $(2\%)$ & \ttt{1.5}{5} \hfill $(1\%)$ & \ttt{2.0}{5} \hfill $(2\%)$ \\
    \raacc   &  \ttt{8.6}{4} \hfill $(96\%)$ & \ttt{4.3}{4} \hfill $(94\%)$ &    \ttt{2.1}{5} \hfill $(79\%)$ & \ttt{9.1}{4} \hfill $(72\%)$ & \ttt{1.2}{5} \hfill $(80\%)$ & \ttt{1.8}{5} \hfill $(93\%)$\\
    \radir   &  \ttt{4.0}{3} \hfill $(4\%)$ & \ttt{1.1}{3} \hfill $(2\%)$ &    \ttt{2.6}{4} \hfill $(10\%)$ & \ttt{8.8}{3} \hfill $(8\%)$ & \ttt{3.2}{3} \hfill $(2\%)$ & \ttt{9.5}{2} \hfill $(1\%)$\\
    \rbdir   &  \ttt{6.5}{2} \hfill $(\lesssim1\%)$ & \ttt{1.8}{4} \hfill $(4\%)$ &    \ttt{2.8}{4} \hfill $(11\%)$ & \ttt{2.5}{4} \hfill $(20\%)$ & \ttt{2.7}{4} \hfill $(18\%)$ & \ttt{1.3}{4} \hfill $(6\%)$\\
    \rlmxb   &  \ttt{1.6}{4} \hfill $(18\%)$ & \ttt{6.5}{3} \hfill $(14\%)$ &    \ttt{1.2}{4} \hfill $(4\%)$ & \ttt{6.0}{3} \hfill $(5\%)$ & \ttt{7.0}{3} \hfill $(5\%)$ & \ttt{5.8}{3} \hfill $(3\%)$\\
    \rdqs   &  \dash & \dash &    \ttt{4.2}{3} \hfill $(8\%)$ & \ttt{3.5}{3} \hfill $(1\%)$ & \dash & \dash
    \enddata
    \tablecomments{The table shows present numbers of QSs per MWEG for models
        with different limiting mass (\mhmax) and different metallicities.
        Results are shown both for the entire population and specific
        evolutionary routes. Numbers in parenthesis represent: for \nqs:
        fraction of all compact objects (NS or QS; \fqs); evolutionary routes:
        fraction of \nqs; \rdqs: fraction of all double compact objects (NS or
        QS). See Sec.~\ref{sec:mhmax} for discussion.}
    \label{tab:comparison}
\end{deluxetable*}

We found that our results change only quantitatively with different values of
\mhmax. The formation of QSs occurs through the same evolutionary routes and the
differences in the coexistence range, which for $\mhmax=1.6\msun$ is between
$1.46\dash1.6\msun$, are also small ($30\dash72\%$). Tab.~\ref{tab:comparison}
provides results for the two values of $\mhmax=1.5$ or $1.6\msun$. The \nqs\
changes by a factor of $\sim2$. The \raacc\ is the main evolutionary route for
both models with more than $72\%$ of QSs forming through this scenario. The
fraction of LMXBs is similar for both values of \mhmax\ (difference of
$\lesssim4\%$). 

A significant difference was obtained for the number of DQS, which dropped by
nearly two orders of magnitude. The higher value of \mhmax\ requires higher
initial stellar masses, as both QSs form directly after the SN (not as a result
of mass accretion). Massive stars are less numerous on ZAMS, therefore DQSs have
less progenitors for higher \mhmax.

\section{Discussion}

\subsection{A comparison with a previous study}

\citet{Belczynski0203} performed a population synthesis study of a QS population
with the use of the earlier version of the \startrack\ code. They found that QSs
may constitute $\sim10\%$ of all compact objects and noted that most of them in
the Galaxy will be single rather than bound with companions. The current version
of the \startrack\ code has been significantly updated since that paper (see
Sec.~\ref{sec:modeling}). Moreover, we incorporated a much more realistic model
of QSs based on the two-families scenario. \citet{Belczynski0203} just assumed
that a fraction of stars in a particular mass range represent QSs. 

Nevertheless, our results mostly agree with those of \citet{Belczynski0203}. We
also found that the majority of QSs in the MW galaxy exist as single stars and
that their number, although being significantly smaller than the number of NSs,
is comparable with the number of BHs. The present study is also broader as it
involves additionally an analysis of the formation scenarios, DQS mergers, and
LMXBs.

\subsection{Comparison with observations}

The deconfinement process modifies the mass distribution of compact stars in the
coexistence range.  (Figs.~\ref{fig:qsmasses_all} and \ref{fig:qsmasses_lmxb}).
Our calculations predict $10\dash57\%$ more binaries in models involving
deconfinement than in models without it (noQS). For LMXBs this excess is even
more pronounced ($22\dash61\%$). For $\mhmax=1.6$ the excess is larger:
$57\dash72\%$ for all binaries, and $30\dash67\%$ for LMXBs, but corresponds to
a less populated range of masses and, therefore, will be even harder to detect.
Nevertheless, the peak of the compact object mass distribution is located
outside of the coexistence range, thus this range is a less populated part of
the distribution. Consequently, current statistics of NS mass measurements
\citep[e.g.][]{Ritter0306,Lattimer1211,Ozel1609} are too small to prove or
reject the presence of this excess. However, even just by having reliable mass
distributions for compact objects, for example, in LMXBs or for millisecond
radio pulsars \citep{Kiziltan1311}, one can try to study if some features are
related to the presence of QSs in the population.

According to our results, most of the QSs are accompanied by WDs
($78\dash95\%$). On average it is also true for LMXBs, but it is more
model-dependent ($7\dash84\%$). In general, the fraction of WDs is greater for
higher metallicities, but seems to be independent on \mhmax. Even if a QS formed
with a RG companion (\raacc), the counterpart will mostly become a WD at some
age. Therefore, QSs will spend typically most of theirs life with a WD
companion. As far as observations are concerned, \citet{Lattimer1211} provided a
list of NS mass measurements in binaries and the most typical companions
appeared to be WDs. What makes WD the most typical companion is the long
duration of this evolutionary stage. Therefore, we have higher chance of
observing the system in that time. Although WDs are significantly lighter than
QSs during MT, the resulting orbit expansion is counteracted by WD expansion and
GR, what allows for a prolonged MT. It is easier to fill the RL by a companion
which expands significantly due to nuclear evolution (e.g. MS, RG), however, a
WD, if it manages to fill its RL, will provide a much longer MT phase. 

In the near future more accreting compact objects can be identified in an X-ray
survey made by eROSITA \citep{Predehl1007} on-board Spectrum-RG satellite (to be
launched in 2018). Systems with WD donors are of special interest, as in such
cases accretors are expected to be massive. Accordingly, our simulations predict
that WDs should be the most typical donors to LMXBs with QSs. After accretion is
over a compact object can be observed as a millisecond radio pulsar. It is
expected that the new radio telescope FAST \citep{Nan11} can provide more
sources of this kind.

\subsection{Phenomenology of QSs in LMXBs}

As one can notice from Tab.~\ref{tab:comparison}, the fraction of LMXBs
containing a QS is not negligible, ranging from a few percent for low
metallicities to almost $20\%$ for solar metallicity.  We estimated the rate of
formation of QSs in LMXBs to be 19.5 (12.7) / 23.9 (16.7) / 15.8 (11.6) events
$\myr^{-1}$ MWEG$^{-1}$ for \zsun\ / $10\%$ of \zsun\ / $1\%$ of \zsun,
respectively (numbers in parenthesis refer to model with $\mhmax=1.6\msun$).
Assuming that Milky-Way (MW) galaxy consists of 1/6 Population II stars and 5/6
Population I, we get an estimated number of $\sim13\dash20$ events $\myr^{-1}$
in MW connected with NS to QS transition.

There are at least two possible observational implications of this result: the
emission of a powerful electromagnetic signal in correspondence with the
formation of a QS and the spin distribution of the pulsars in LMXBs. 

The formation of a QS in a LMXB is a strongly exothermic process (releasing
order of $10^{53}$ erg) and it can take place in a millisecond radio pulsar.
These two properties strongly suggest a connection between the formation of a QS
in a LMXB and at least a sub-class of GRBs within the protomagnetar model
\cite{Metzger:2010pp}.  It is remarkable that such a GRB would not be connected
with the death of a massive star and thus with a SN.  It is tempting to
associate this possibility with the famous case of GRB060614
\citep{Fynbo:2006mw,DellaValle:2006nf,GalYam:2006en}.

Although it is difficult to derive a frequency from just a single event we can
try to compare the observed "rate" of GRBs lacking a SN with the rate of GRBs
associated with the formation of a QS in a LMXB (mainly route $\raacc$).

\begin{itemize}
    \item The rate of NS to QS transitions in LMXBs is of the order of
        $\sim13\dash20$ events $\myr^{-1}$. 

    \item A significant fraction, order of few tens percent, of compact stars in
        LMXBs rotates very rapidly and could possibly generate a GRB (always
        through the protomagnetar mechanism). This translates into a rate of
        GRBs associated with $\rlmxb$ of the order of one every $10^6$ years.

    \item The fraction of long GRBs lacking a SN in respect to the GRBs for
        which an association with a SN has been clearly established to be of the
        order of $10\%$ \citep{Hjorth:2011zx}.

    \item The rate of long GRBs has been estimated to be of the order of one
        every $10^5-10^6$ years per galaxy\citep{Podsiadlowski:2004mt},
        therefore the rate of long GRBs non associated with a SN could be of the
        order of one every $10^6-10^7$ years. One can notice that the rate
        estimated in our model is fairly close to the observed one. 
\end{itemize}

One should note that within the protomagnetar model a very strong magnetic field
is needed. If this magnetic field is present before the formation of the QS, it
may hinder the mass accretion. The magnetic field could instead be generated
during the combustion from hadrons to quarks, which lasts a few seconds
\citet{Drago:2015fpa}. During the combustion the moment of inertia increases
significantly \citep{Pili:2016hqo} and it leads to the development of a strong
differential rotation which in turn could generate the needed high magnetic
field \citep{new-bucciantini}.

The second possible phenomenological implication concerns the spin distribution
of fast rotating pulsars in LMXBs. The increase of the moment of inertia
resulting from the conversion of a NS into a QS implies a significant spin-down
of the pulsar. In \citet{Pili:2016hqo} a change of the moment of inertia was
large, up to a factor of two, implying a reduction of the spin frequency again
by a factor of two. It is tempting to connect this effect with the bimodal
distribution of the spin frequency found recently by  \citet{Patruno:2017oum}
where the slowest component would contain a significant fraction of QSs in our
scheme.

\subsection{Strangelets pollution}

The rate of mergers of DQS is crucial in order to estimate the production of
strangelets, i.e. of lumps of stable strange quark matter, significantly smaller
than a star.  There are two known mechanisms by which strangelets could be
produced: they could be produced at the time of primordial baryogenesis (when
the temperature did fall below about 150 MeV) or they could be produced by
partial fragmentation of at least one of the QSs at the beginning of the merging
process of a DQS system \footnote{A further possible mechanism for strangelets
    production would be connected with an explosive conversion of hadronic stars into QSs
    \cite{Jaikumar:2006qx} but we follow the scheme presented in
    Sec.~\ref{sec:modeling} and supported by the papers there quoted in which
    detonation is never obtained.}. 
The first process is uncertain (it has been criticized e.g. in
\citet{Alcock:1985vc}), but the second is very relevant as a potential source of
strangelets. The existence of a significant flux of strangelets could trigger
deconfinement in all compact stars at the moment of their formation
\citep{Madsen:1989pg}, implying that only QSs can exist and therefore
invalidating the two-families scenario. In order to clarify this issue two
crucial information have to be provided: the rate of DQS mergers not directly
collapsing into a BH and the probability of forming fragments (strangelets) in
the mass range indicated above. An estimate of the first number has been
obtained in this simulation. First, the number of DQS mergers is about 12
Gyr$^{-1}$ in our Galaxy, as stated above. Second, the total mass of the binary
system exceeds $3 M_\odot$ in most of the cases. Due to that many of these DQS
systems collapse directly to a BH, as indicated by the analysis of
\citet{Bauswein:2008gx}. The exact fraction of events in which the BH is not
promptly formed is linked to $M_{\mathrm{max}}^Q$, but in any case it cannot
exceed 12 events Gyr$^{-1}$.

Assuming, as an upper limit, that each event releases a mass of about
$10^{-2}M_{\odot}$ (similar to the mass ejected in double NS mergers; the real
number could be smaller by three/four orders of magnitude) one obtains an
average strange quark matter density $\rho_s$ in the Galaxy of about
($10^{-35}-10^{-36}$) g~cm$^{-3}$. The flux of strangelets per unit of solid
angle $\mathrm{d}j_s/{\mathrm{d}\Omega}$ can be estimated as follows by assuming
that they all have the same baryon number A: 

\begin{equation}
    \frac{\mathrm{d}j_s}{\mathrm{d}\Omega}=\frac{\rho_s v}{4 \pi A m_p},
\end{equation} 

\noindent where $v$ is the average velocity of the strangelets and $m_p$ is
the proton mass.  By assuming that low mass strangelets have a velocity
comparable to the velocity of the galactic halo i.e.  $v=250$~km s$^{-1}$ one
gets: ${\mathrm{d}j_s}/{\mathrm{d}\Omega} \sim {10^{-5}}
\rho_{35}/A$~cm$^{-2}$~s$^{-1}$ sr$^{-1}$ where
$\rho_{35}={\rho_s}/({10^{-35}~\mathrm{g/cm^3}}).$   

Having estimated an upper limit to the flux of strangelets it is possible to
compare this limit with limits coming from Earth and Lunar experiments and with
limits coming from astrophysics.  Concerning the first type of limits,
summarized in
\citet{Price:1983ax,DeRujula:1984axn,PerilloIsaac:1998xm,Weber:2004kj,Han:2009sj},
they are almost completely respected by our estimate of the flux. Only taking
our very conservative upper limit on $\rho_s$, a small overlap with the
constraints from the Lunar Soil experiment is found. 

A more stringent constraint has been obtained recently by the PAMELA experiment
\citep{Adriani:2015epa}. Our upper limit on the flux would violate the
observational limits for $A \lesssim 10^3$. On the other hand, the more
realistic estimate quoted above fully satisfies the PAMELA limits.

Concerning the limits coming from astrophysics, the most relevant analysis has
been done by \citet{Madsen:1989pg}.  First, even using our highest value of
$\rho_s$ the probability of capture of strangelets by a cold NS is negligible.
This implies that pulsars displaying glitches (such as the Vela and the Crab
pulsars) had a marginal chance to transform into QSs (which could not be able to
glitch). In this way one of main mechanisms for the conversion of all NSs into
QSs is ruled out. 

According to Madsen, another possibility to trigger the formation of a QS is
based on the capture of strangelets by main sequence stars: the strangelets
would accumulate close to the core of the star and they would transform a NS
into a QS soon after the SN explosion. In order to be captured by a main
sequence star (and not to pass through it) strangelets need to have a baryon
number smaller than $\sim 10^{28}$.  This mechanism has two weak points. First,
it is easy to demonstrate by using dimensional arguments that the strangelets
dynamically ejected at the moment of the merger have a baryon number larger than
about $10^{38}$ \citep{Madsen:2001bw}. Strangelet fragmentation through
collisions, while quite efficient, could not be able to reduce the baryon
number of the strangelets initially produced by ten orders of magnitude
\citep{new-bucciantini}. Second, the strangelet located in the core of the
collapsing star could evaporate due to the high temperatures reached at the
moment of the bounce.  Similar arguments can be applied to the case of molten
NSs. 

In conclusion, the limits stemming from Earth and Lunar experiments can be
rather easily satisfied directly by our estimate of the upper limit on the flux
without making any assumption on the fragmentation mechanism of the strangelets.
Astrophysical limits are more subtle: in particular they depend on the ability
of the strangelets to fragment into small nuggets and to survive temperatures of
the order of few MeV.     

\subsection{Single QS population}

Results of presented simulations show that formation of a QS in a binary system
usually results in a disruption.  Only in $\sim3\dash10\%$ of cases the stars
remain bound.  As a result, about $(2.3\dash2.4)\times10^6$ single QSs
(depending on the model) originating from disrupted binaries should be present
currently in a MWEG. Potentially it is possible to form a QS through single-star
evolution providing the ZAMS mass of a star is in the range
$\mzams\approx17\dash22\msun$. However, most often so massive star are found in
binaries \citep[e.g. binary fraction $>90\%$ found by][]{Sana1411}.

We note that a QS may be a result of a merger of a DQS.  The number of such
events is very low, as we show in Sec.~\ref{sec:dqs}.  We do not consider this
possibility here.  Also, a few single QSs can result from NS-NS coalescence.
Double NS coalescences have a rate of about one in $10\dash20 \kyr$ in a MWEG
(\citet{Postnov:2014py}).  If in a few percent of such collisions stable QSs are
formed (Drago et al., in prep.), then we have about a few tens of thousands of
isolated QSs formed via this channel in a galactic lifetime.

\section{Summary and Conclusions}

We performed a population synthesis study of strange quark stars (QS). The two
families scenario predicts that a neutron star (NS) becomes a QS after reaching
the mass limit  $\mhmax$ \citep{Drago1602}, which we adopted to be $1.5$, or
$1.6\msun$ in our modeling.  Our results turn out to be rather robust respect to
the variation of $\mhmax$. Notice anyway that in our analysis we have not
included the effect of rapid rotation on the structure of the star. This will
constitute the next extension of the present work.

Our analysis of QS population may be summarized as follows:

\begin{itemize}

\item We found that QS may constitute $\sim 1\dash4\%$ of all compact objects in
    binaries (moreover, in our scheme all compact objects with masses larger
    than $\sim 1.5-1.6\msun$ are QSs). Typically, a QS forms as a result of
    mass-accretion from a red giant companion onto a NS, however, a direct
    formation (immediately after the supernova explosion) is also possible in
    $\lesssim30\%$ of cases.

\item A relatively larger number of QS is predicted in low-mass X-ray binaries
    ($3\dash18\%$) and especially in the coexistence range ($22\dash72\%$). The
    effect on the mass distribution of compact stars is, however, too small to
    be detected using current observations. If future missions will provide
    better mass and radius measurements, it will be possible to test our
    predictions.

\item Double QSs may constitute up to $8\%$ of double compact objects with
    components masses below $2.5\msun$. In most of the cases the two QSs do not
    merge within a Hubble time. We estimated a merger rate of $\sim12$ events
    $\gyr^{-1}$ for the Galactic bulge. Such a low rate implies a rather small
    ``strangelets pollution'' and in turn rules out at least one of the possible
    mechanisms suggested in the literature to convert all NSs into QSs.
    Moreover, all limits stemming from Earth and Lunar experiments are rather
    easily satisfied.

\item The rate of conversion of a NS into a QS due to mass accretion in low-mass
    X-ray binaries is rather large, order of one event every $10^6$ years. This
    process is strongly exothermic (it releases about $10^{53}$ erg) and it can
    take place in a rapid rotating compact star. These two properties suggest
    a possible connection with a special subclass of gamma-ray-bursts, possibly
    including GRB060614.

\end{itemize}

\acknowledgements

We want to thank thousands of volunteers who supported the research by
participation in Universe@Home project\footnote{http://universeathome.pl} and
the anonymous referee for valuable comments.  GW was partially supported by
Polish NCN grant No. UMO-2015/19/B/ST9/03188. PS acknowledges support from the
Russian Science Foundation grant 14-12-00146.  

\bibliographystyle{apj}
\bibliography{ms}

\end{document}